\documentclass[aps,prl,
twocolumn
]{revtex4-1}
\usepackage{amssymb}
\usepackage{amsmath}
\usepackage{graphicx}
\usepackage{xfrac} 
\usepackage{bm}
\usepackage
{todonotes} 

\begin{document}
 \title{On the optimal calculation of the pair correlation function for an orthorombic system}
 \date{\today}
 \author{Kai A. F. R\"ohrig}
 \affiliation{Institute of Physics, Johannes Gutenberg University Mainz, Staudinger Weg 7, D-55128 Mainz}
 \author{Thomas D. K\"uhne}
 \affiliation{Institute of Physical Chemistry and Center of Computational Sciences, Johannes Gutenberg University Mainz, Staudinger Weg 7, D-55128 Mainz}
 
\begin{abstract}
We present a new computational method to calculate arbitrary pair correlation functions of an orthorombic system in the most efficient way. The algorithm is demonstrated by the calculation of the radial distribution function of shock compressed liquid hydrogen. 
\end{abstract}
 
\maketitle
 
Determining the equilibrium properties of a classical many-body system is rather challenging due to the large variety of behavior, which results from the interactions between the particles. 
The atomic structure of a system is best characterized by a set of correlation functions 
\begin{eqnarray}
  g^{(n)}(\bm{R}_{1}, ..., \bm{R}_{n}) &=& \frac{V^{n}N! Z_{N}^{-1}}{N^{n} (N-n)!} \int_{D(V)}{d\bm{R}_{n+1} \cdot \cdot \cdot d\bm{R}_{N}}  \nonumber \\
  &\times& e^{- \beta U_{N}(\bm{R}_{1}, ..., \bm{R}_{N})} \label{PCF}, 
\end{eqnarray}
where $D(V)$ is the spatial domain defined by the physical container, $N$ the number of particles, $V$ the volume, $\beta=1/k_{B}T$ the inverse temperature and $U_{N}(\bm{R}_{1}, ..., \bm{R}_{n})$ the potential energy function of the system, while $n$ is the number of fixed particles. The simplest is the so called pair-correlation function (PCF) $g^{(2)}(\bm{R}_{1}, \bm{R}_{2}) \equiv g(\bm{r})$, which is of particular importance, as it is related to the structure factor $S(\bm{q}) = 1 + \rho \int_{D(V)}{d\bm{r} \, e^{-i \bm{q}\bm{r}} [g(\bm{r}) - 1]}$ by the Fourier transform. For a spatially isotropic system, such as a liquid, the PCF does only depend on the absolute value of the relative distance $|\bm{r}|=r$ and can be simplified to the radial distribution function (RDF) $g(r)$, which gives the probability of finding a pair of atoms a distance $r$ apart, relative to the probability of a ideal gas at the same density. Like before, the Fourier transform of $g(r)$ corresponds to the isotropic structure factor $S(|\bm{q}|) = 1 + \rho \int_{D(V)}{d\bm{r} \, e^{-i \bm{q}\bm{r}} [g(r) - 1]}$, which can be experimentally measured using neutron scattering or X-ray diffraction techniques \cite{Pedersen1997, Soper2007}. In addition, it is not only possible to calculate the potential of mean force, but assuming that $U_{N}(\bm{R}_{1}, ..., \bm{R}_{n})$ is a pair-potential, also numerous important thermodynamic properties such as the potential energy and the pressure, just to name a few.  

In order to compute the RDF in a periodic cubic cell of length $L$, it is customary to restrict oneself to consider only those atoms that lie within the sphere of radius $L/2$ and discard all the rest. In this most favorable case, the simulation cell needs to be $3\sqrt{3} \approx 5.2$ times larger than would be in principal necessary to extract the same information. Particularly, in the context of \textit{ab-initio} molecular dynamics (MD) \cite{CarParrinello1985, Kuehne2007, Kuehne2009} this results in a substantial additional computational burden. The situation is even worse for a general orthorhomic unit cell, where the fraction of the exploited volume may become arbitrarily small. For quasi one-dimensional systems, such as nanowires or nanotubes, this does indeed occur, just as in slab calculations of two-dimensional surfaces, or during shock compression simulations, where the unit cell is either rather prolate or oblate. 

For the simplest case of a cube in three dimensions, Deserno has shown how to extend the permissible radius to the maximum value given by the particular unit cell \cite{Deserno}. In this Brief Report, we will generalize this approach to arbitrary orthorhombic unit cells with lattice parameters $(a,b,c)$. 

Let us start be rewriting the PCF in terms of $\delta$-functions for the first $n$ particles and integrating over all $N$ particles: 
\begin{subequations}
\begin{eqnarray}
  g^{(n)}(\bm{R}_{1}, ..., \bm{R}_{n}) &=& \frac{V^{n}N! Z_{N}^{-1}}{N^{n} (N-n)!} \int_{D(V)}{d\bm{R}'_{1} \cdot \cdot \cdot d\bm{R}'_{N}} \nonumber \\
  &\times& e^{- \beta U_{N}(\bm{R}'_{1}, ..., \bm{R}'_{N})} \prod_{i=1}^{n}{\delta(\bm{R}_{i} - \bm{R}'_{i})} \label{DeltaVal} \\
  &=& \frac{V^{n}N!}{N^{n} (N-n)!} \left< \prod_{i=1}^{n}{\delta(\bm{R}_{i} - \bm{R}'_{i})} \right>_{\bm{R}'_{1}, ..., \bm{R}'_{N}}
  \label{ExpVal}
\end{eqnarray}
\end{subequations}
The latter denotes the ensemble average of the quantity $\prod_{i=1}^{n}{\delta(\bm{R}_{i} - \bm{R}'_{i})}$, using $\bm{R}'_{1}, ..., \bm{R}'_{N}$ as integration variables. Again, confining ourself to the most relevant case $n=2$ and invoking isotropy, Eq.~(\ref{ExpVal}) can be written as
\begin{eqnarray}
  g(r) &=& \frac{N(N-1)}{\rho^{2}} \frac{\int_{D(V)} {d\bm{R}'_{1} \cdot \cdot \cdot d\bm{R}'_{N}} \, e^{- \beta U_{N}(\bm{R}'_{1}, ..., \bm{R}'_{N})} \, G (r)}{\int_{D(V)} {d\bm{R}'_{1} \cdot \cdot \cdot d\bm{R}'_{N} \, e^{- \beta U_{N}(\bm{R}'_{1}, ..., \bm{R}'_{N})}}} \nonumber \\
  &=& \frac{\int_{D(V)} {d\bm{R}'_{1} \cdot \cdot \cdot d\bm{R}'_{N}} \, e^{- \beta U_{N}(\bm{R}'_{1}, ..., \bm{R}'_{N})} \, G (r)}{\rho^{2} N^{-2}(1-N)\int_{D(V)} {d\bm{R}'_{1} \cdot \cdot \cdot d\bm{R}'_{N} \, e^{- \beta U_{N}(\bm{R}'_{1}, ..., \bm{R}'_{N})}}} \nonumber \\
  &:=& \frac{\tilde G(r)}{A(r)}, \label{EnsAvg}
\end{eqnarray}
where $\rho=N/V$ is the particle density, while $G(r)$ may be any arbitrary quantity for which to calculate the RDF. In a numerical calculation, the numerator $\tilde G(r)$ is computed by averaging $G(r)$ over shells of constant radius via a binning procedure, while the denominator $A(r)$ plays the role of a normalization and depends only on the geometry of the simulation box. However, due to the discretization of space, we cannot directly compute $\tilde G(r)$, but solely its integrated value $F(r)$ over the length of a bin $\Delta r$:
\begin{eqnarray}
  \int_{r}^{r+\Delta r} dr^\prime \, \tilde G (r^\prime) :=  F(r ) \label{IntNom}
\end{eqnarray}
The choice of normalization for $F(r)$ is arbitrary, except for the constraint that the normalization must reproduce Eq.~(\ref{EnsAvg}), i.e. in the limit $\Delta r \to 0$ it converges to the exact solution. The most straightforward choice of $A(r)$ is 
\begin{eqnarray}
  \int_{r}^{r+\Delta r} dr^\prime \, A (r^\prime) =  \tilde{V} ( r + \Delta r ) - \tilde{V} (r), \label{DenNom} 
\end{eqnarray}
where $\tilde{V}(r)  := \int_{0}^{r} dr^\prime \, A (r^\prime)$ is the volume of the sphere that is intersecting with the simulation cell. Instead of Eq.~(\ref{DenNom}), $A(r) \times \Delta r$ can be employed as an alternative normalization. The fact that both are indeed genuine normalizations can be seen using the rule of l'H\^{o}pital: 
\begin{eqnarray}
  \hspace{-0.5cm} \lim_{\Delta r \to 0} \frac{F (r)}{ \tilde{V} ( r + \Delta r  ) - \tilde{V} ( r )  } = \lim_{\Delta r \to 0} \frac{F (r)}{ A(r) \times \Delta r } = \frac{\tilde G ( r)}{ A ( r)} \label{lHopital}
\end{eqnarray}

Since $F(r)$ can be directly extracted from a MD or Monte Carlo simulation, in the following we will only discuss how to evaluate Eq.~(\ref{DenNom}). To that extent, we will provide an explicit set of formulae to compute $\tilde{V}(r)$ for the general case of an orthorhombic simulation cell.
 
If the size of the simulation cell is infinite, the normalization function $\tilde{V}(r)$ would simply correspond to the volume of a sphere, i.e. $\tilde{V}_{0}(r) = \frac{4}{3} \pi r^3$. For a finite simulation cell though, $\tilde{V}(r)$ is the intersecting volume of the sphere with the simulation box. This requires a distinction of cases for the different types of intersections, which complicates matters. 
The simplest solution is to circumvent this difficulty by confining $r$, so that $0 < 2r \leq \min(L_x , L_y , L_z)$ holds. In this way $\tilde{V}_{0}(r) = \frac{4}{3} \pi r^3$ is still valid, but unfortunately, any additional information outside the sphere is neglected. Thus, depending on the shape of the simulation cell, the utilized radius is at least $\sqrt{3} \approx 1.7$ times smaller than ideally possible. However, the latter requires to compute $\tilde{V}(r)$ for all different types of intersections. For a general orthorhombic simulation cell with dimensions $a \ge b \ge c$, we can distinguish between seven critical values of $r$: 
\begin{eqnarray}
 \frac{c}{2} &\le& \frac{b}{2} \le \frac{a}{2} \le \frac{1}{2}  \sqrt{c^2 + b^2} \le \frac{1}{2} \sqrt{c^2 + a^2} \nonumber \\ 
 &\le& \frac{1}{2} \sqrt{b^2 + a^2} \le \frac{1}{2} \sqrt{c^2 + b^2 + a^2} \label{CriticalR}
\end{eqnarray}
 
This results in the following eight intervals for $r$ and corresponding volume functions, which are denoted as $I$-$VIII$ and $\tilde{V}_I (r)$~through~$\tilde{V}_{VIII}(r)$, respectively: 
\begin{eqnarray*}
  I &:& \qquad  0 < 2r \le c \\
  II &:& \qquad c < 2r \le b \\
  III &:& \qquad b < 2r \le \min(a,\sqrt{c^2 + b^2}) \\
  IV &:& \qquad \min(a,\sqrt{c^2 + b^2}) < 2r \le  \max(a,\sqrt{c^2 + b^2}) \\
  &\,& \qquad IV_a : a \le \sqrt{c^2 + b^2} \; , ~~ IV_b :  \sqrt{c^2 + b^2} < a \\
  V &:& \qquad \max(a,\sqrt{c^2 + b^2}) < 2r \le  \sqrt{c^2 + a^2} \\
  VI &:& \qquad \sqrt{c^2 + a^2} < 2r \le  \sqrt{b^2 + a^2} \\
  VII &:& \qquad \sqrt{b^2 + a^2} < 2r \le  \sqrt{c^2 + b^2 + a^2} \\
  VIII &:& \qquad \sqrt{c^2 + b^2 + a^2} <  2r 
\end{eqnarray*}
 
The cases $IV_a$ and $IV_b$ are mutually exclusive, and it depends on the cell geometry, which one is appropriate. In cartesian coordinates, $\tilde{V}(r)$ reads as 
\begin{eqnarray}
  \tilde{V} ( r) = 8 \int_{0}^{X(r)}  dx  \int_{0}^{Y(r,x)}  dy \int_{0}^{Z(r,x,y)}  dz, \label{NewVol} 
\end{eqnarray}
where the integration limits are 
\begin{subequations}
\begin{eqnarray}
  X(r) &=& \min \left(r, \frac{c}{2} \right), \\
  Y(r, x) &=& \min \left(\sqrt{r^2 - x^2}, \frac{b}{2} \right)~\text{and} \\
  Z(r, x, y) &=& \min \left(\sqrt{r^2 - x^2 - y^2}, \frac{a}{2} \right), 
\end{eqnarray}
\end{subequations}
respectively. 

Solving Eq.~(\ref{NewVol}) analytically for each of the aforementioned cases, eventually leads to 
\begin{eqnarray*}
  \tilde{V}_{I}(r) &=& \frac{4}{3} \pi r^3 \\
  \tilde{V}_{II}(r) &=& V_{I} - 2 r^3 \times V_K (c / 2r) \\
  \tilde{V}_{III}(r) &=& V_{II} - 2 r^3 \times V_K (b / 2r) \\
  \tilde{V}_{IV_a}(r) &=& V_{III} - 2 r^3 \times V_K (a / 2r) \\
  \tilde{V}_{IV_b}(r) &=& V_{III} + 4 r^3 \times V_N(c / 2r, b / 2r ) \\
  \tilde{V}_{V}(r) &=& V_{III} - 2 r^3 \times V_K ( a / 2r ) \\
  &+& 4 r^3 \times V_N ( c / 2r, b / 2r ) \\
  \tilde{V}_{VI}(r) &=& V_{V} + 4 r^3 \times V_N (c / 2r, a / 2r ) \\
  \tilde{V}_{VII}(r) &=& V_{VI} + 4 r^3 \times V_N ( b / 2r, a / 2r ) \\
  \tilde{V}_{VIII}(r) &=& a b c, 
\end{eqnarray*}
where we have introduced the shorthand notations
\begin{subequations}
\begin{eqnarray}
  V_K ( \alpha ) &:=&  \pi \left[ \frac{2}{3} - \alpha +\frac{1}{3} \alpha ^3   \right]~\text{and} \label{VolSphericDome} \\ 
  V_N (\beta, \gamma) &:=& \frac{2}{3} \left[ H( \beta , \gamma ) + H( \gamma , \beta ) \right],~\text{with} \label{VolWedgeSphere} \\ 
  H(\omega, \tau)&:=&  \int^{ atan(\omega / \tau ) }_{asin(\omega )} d \phi \left[ 1- \left( \frac{\omega}{sin(\phi)} \right)^{2} \right]^{3/2} \\
  &=& atan \left (\frac{\omega}{\tau} \sqrt{1 - \omega^2 - \tau^2} \right ) \nonumber \\
  &+& \frac{3 \omega - \omega ^3}{2} \left[ atan \left(\frac{\tau}{\sqrt{1 - \omega^2 - \tau^2}} \right)  - \pi / 2 \right] \nonumber \\
  &+& \frac{\omega \tau}{2} \sqrt{1 - \omega^2 - \tau^2}.
\end{eqnarray}
\end{subequations}
Herein, $V_K (\alpha)$ denotes the volume of a spheric dome of radius one that is cutoff at a height $z=\alpha$, 
whereas $V_N (\beta, \gamma)$ corresponds to the volume of a wedge cut from the unit sphere, whose planar sides lie at $x=\beta$ and $y=\gamma$. Since the auxiliary functions $V_K(\alpha)$ and $H(\omega, \tau)$ are only needed for $\alpha \in \left[0,1\right]$ and $\omega^2 + \tau^2 \in \left[0,1 \right]$, respectively, it is possible to exploit this in the interest of an efficient implementation. 

\begin{figure}
 \includegraphics[width=9.5cm]{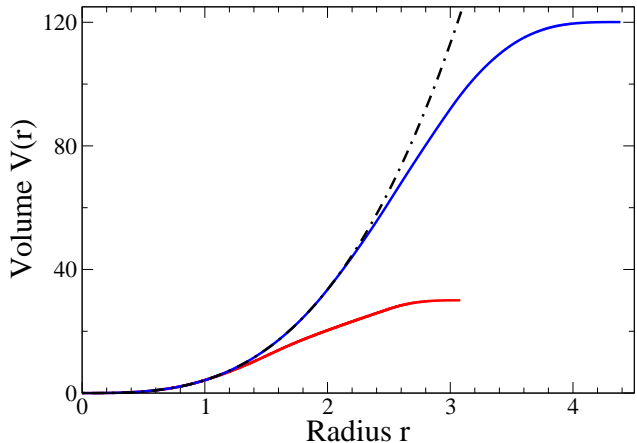}
 \caption{$\tilde{V}(r)$ for an orthorhombic unit cell with dimensions $2\times3\times5$ (red) and $4\times5\times6$ (blue). The dashed line indicates $V_{0}(r)=\frac{4}{3} \pi r^3$.} \label{Vdemo}
\end{figure}

\begin{figure}
 \includegraphics[width=9.5cm]{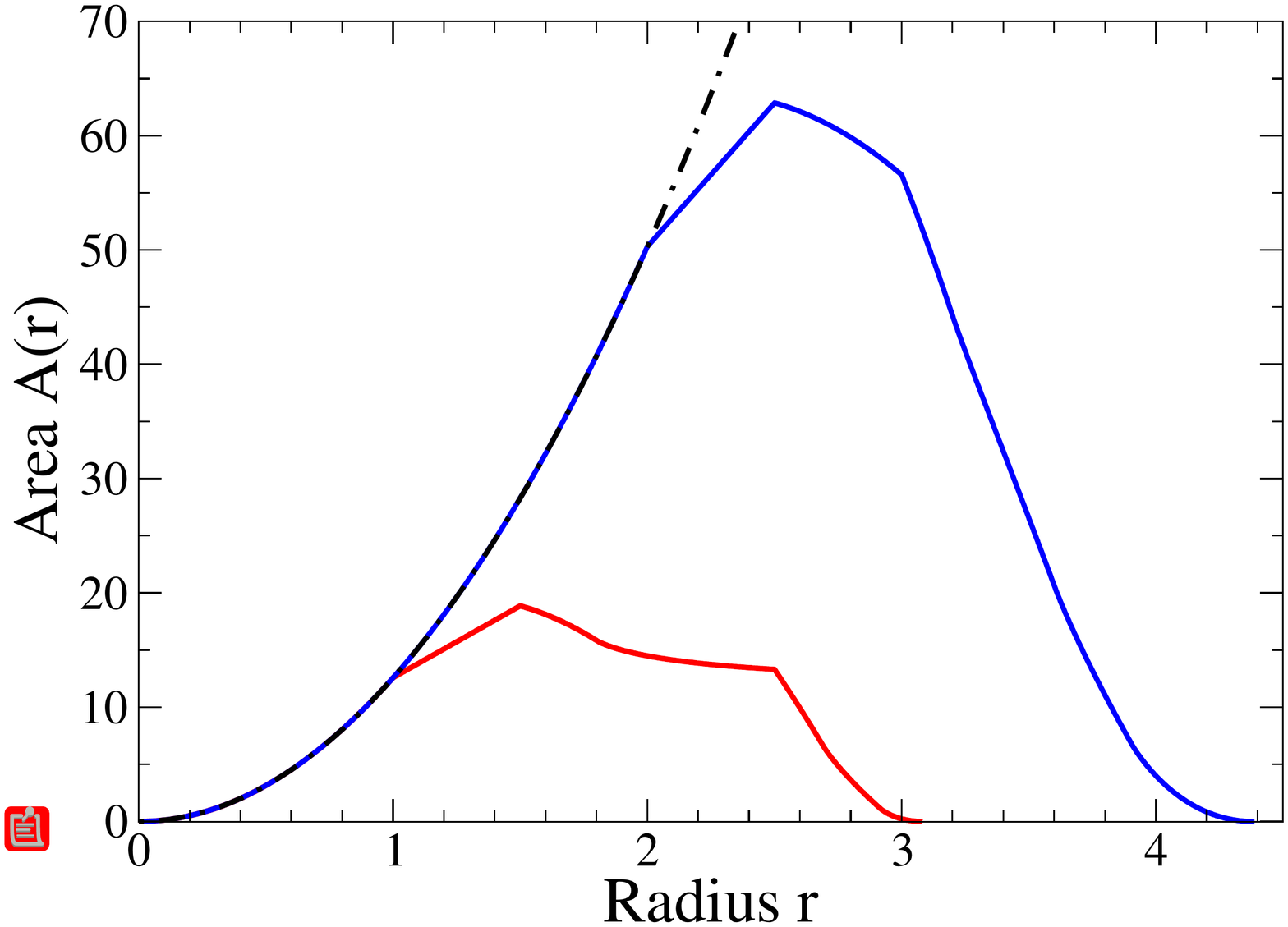}
 \caption{$A(r) = \partial_r \tilde{V}(r)$ for an orthorhombic unit cell with dimensions $2\times3\times5$ (red) and $4\times5\times6$ (blue). The dashed line indicates $A_{0}(r)=4\pi r^2$.} \label{Ademo}
\end{figure}

In Fig.~\ref{Vdemo} and \ref{Ademo} the eventual normalization function $\tilde{V}(r)$ and the denominator $A(r) = \partial_r \tilde{V}(r)$ of Eq.~(\ref{EnsAvg}) are shown for two sample simulation cells and compared to $\tilde{V}_{0}(r)=\frac{4}{3}\pi r^{3}$ and $A_{0}(r)=4\pi r^2$, respectively. As can be directly seen, the accuracy of the functions $\tilde{V}_{0}(r)$ and $A_{0}(r)$ immediately deteriorates for $r \ge \min(L_x , L_y , L_z)/2$. Although, using the present computational method, it is possible to exploit the whole volume up to the maximum permissible radius $r_{\max} = \sqrt{a^2 + b ^2 + c^2}/2$, the statistics will become inferior with increasing radius. Moreover, for $r \rightarrow r_{\max}$ the evaluation of Eq.~(\ref{EnsAvg}) will not only be affected by statistical uncertainties, but also by a small bias due to the fact that in this case the normalization function is vanishing. 

For the purpose of demonstrating our new calculation method, we present here shock compression simulations of liquid hydrogen to illustrate that our method works well for highly prolate and oblate cell geometries. However, instead of simulating a planar shock wave within a large computational box with many atoms, we have employed the multiscale shock-wave simulation technique (MSST) \cite{ReedPRL, Mundy2008}, which is based on MD as well as the Navier-Stokes equations for compressible flow. In this way, the simulation cell follows a Lagrangian point through the shock wave as if the shock were passing over it, instead of directly simulating the shock-wave itself. This allows to perform shock-compression simulations with a sufficiently small number of atoms, so that it became feasible to perform MSST simulations in conjunction with \textit{ab-initio} MD. 

Due to the fact that a direct Born-Oppenheimer MD (BOMD) simulation, where the total energy functional is fully optimized in every MD step, would have been prohibitive, we employ here the recently devised "Car-Parrinello-like approach to Born-Oppenheimer MD" of K\"uhne et al. \cite{Kuehne2007}, which has already demonstrated its superior efficiency \cite{Caravati2007, *Kuehne2011, *Luduena2011a}.  In the spirit of the original Car-Parrinello MD (CPMD) method \cite{CarParrinello1985} during the dynamics, the electronic wave functions are not self-consistently optimized. Nevertheless, at variance to CPMD, in this approach the fictitious Newtonian dynamics of the electrons and ions is substituted by a similarly coupled electron-ion MD, which does not require the definition of a fictitious mass parameter, but at the same time keeps the electrons very close to their instantaneous electronic ground state. As a consequence, the time step can be chosen up to the ionic resonance limit, while simultaneously preserving the efficiency of CPMD. As a consequence, the best aspects of the BOMD and CPMD schemes are unified, which not only extends the scope of either method but allows for \textit{ab-initio} simulations previously thought to be not feasible. 

The MSST simulation has been performed within density functional theory using the mixed Gaussian and plane wave \cite{GPW} code CP2K/Quickstep \cite{Quickstep}, where the electron density is represented by a plane wave basis set, while the orbitals are expanded in Gaussians. Together with efficient transformation methods to switch between one representation or the other and advanced multigrid, screening as well as sparse matrix methods, an efficient linear-scaling evaluation of the Hamilton matrix is obtained.  Efforts towards a full linear-scaling algorithm are underway \cite{Ceriotti2008, *Ceriotti2009}. Here the orbitals are described by an accurate double-$\zeta$ set with one set of polarization functions (DZVP) \cite{MolOptBasis}, whereas a density cutoff of 200~Ry is employed for the charge density. The unknown exchange and correlation potential is substituted by the PBE  generalized gradient approximation \cite{PBE}. The interactions between the valence electrons and the ionic cores are described by rather hard norm-conserving pseudopotentials \cite{GTH, PP}. 

The system consists of 768 hydrogen atoms in an orthorombic box of initial dimensions 68.2~{\AA} $\times$ 11.0~{\AA} $\times$ 11.0~{\AA}. Before starting the actual MSST simulation, the sample has been equilibrated in the NPT ensemble at a temperature of 100~K and a pressure of 1~GPa. Since we are dealing with a rather large disordered system at finite temperature, the Brillouin zone is sampled at the $\Gamma$-point only. The MSST equations of motion for the nuclei and volume of the simulation cell are integrated using a discretized time step of 0.1~fs and a cell mass of $7 \times 10^7$~a.u. ($\widehat{=}$ mass $\times$ length$^{-4}$) to constrain the stress in the shock propagation direction to the Rayleigh line and the energy of the system to the Hugonoit energy condition. 

\begin{figure}
 \includegraphics[width=9.5cm]{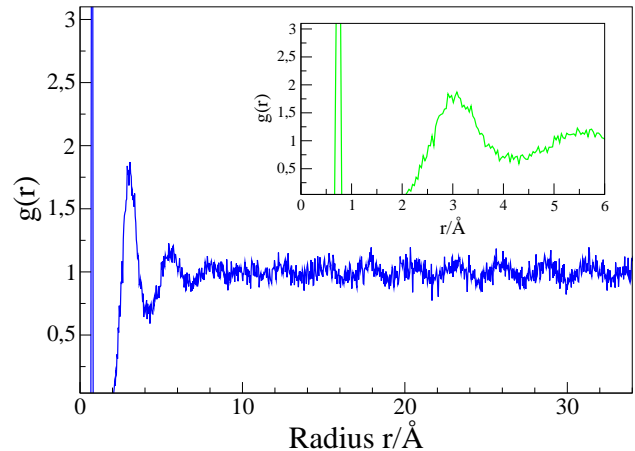}
 \caption{The RDF of liquid molecular Hydrogen before the shock-compression at a temperature of 100~K and a pressure of 1~GPa.} \label{img:example1}
\end{figure}
 
\begin{figure}
 \includegraphics[width=9.5cm]{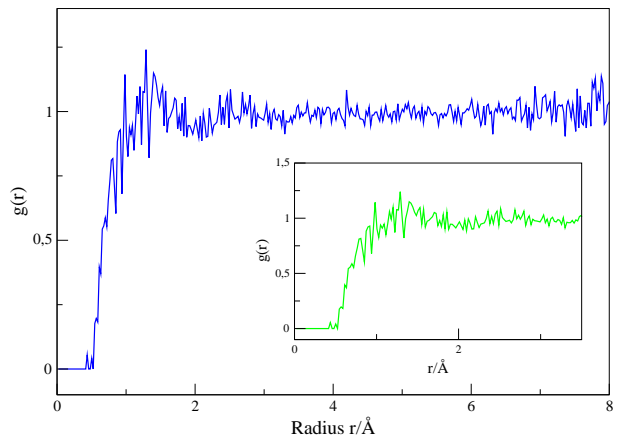}
 \caption{The RDF of liquid atomic Hydrogen after the shock-compression at a temperature of 8500~K and a pressure of 650~GPa.} \label{img:example2}
\end{figure}

Upon applying a shock velocity of 35~km/s along the x-axis, the system undergoes a 9.2-fold compression in volume along the same direction resulting in a simulation box of dimensions 7.4~{\AA}$\times$11.0~{\AA}$\times$11.0~{\AA}. In doing so the pressure increases till 650~GPa, while the temperature raises up to 8500~K. The according RDF's before and after the shock are shown in Fig.~\ref{img:example1} and \ref{img:example2}, which can be compared to the conventional $g(r)$ in the inset. Using the present scheme, it is possible to extract as much information that otherwise would require a simulation cell which is larger by a factor of 38 and 1.5, respectively. In Fig.~ \ref{img:example2} it is also also observed that liquid hydrogen undergoes a molecular-atomic transition to a metal, which is consistent with the findings of Nellis and coworkers \cite{NellisScience, NellisPRL}. However, further details and additional simulations will be published elsewhere. 

We conclude by noting that the observed increase in terms of a accessible volume does neither mitigate single particle finite size effects, nor does it substitute a proper finite size scaling \cite{Binder1981}.

The authors would like to acknowledge financial support from the Graduate School of Excellence MAINZ and the IDEE project of the Carl-Zeiss Foundation.

 
\end{document}